\documentclass[fleqn,usenatbib,usedcolumn]{mnras}

\usepackage{graphicx} 

\usepackage{newtxtext,newtxmath}

\usepackage[T1]{fontenc}

\usepackage{advdate}

\DeclareRobustCommand{\VAN}[3]{#2}
\let\VANthebibliography\thebibliography
\def\thebibliography{\DeclareRobustCommand{\VAN}[3]{##3}\VANthebibliography}

\usepackage{graphicx}	
\usepackage{amsmath}	

\usepackage{amssymb}	

\graphicspath{{./}{./figures/}} 

\title[Diffusion and back-reaction effects on dust]{%
Effects of turbulent diffusion and back-reaction on the dust distribution around two resonant planets}

\author[Marzari and D'Angelo]{
Francesco Marzari$^{1}$\thanks{E-mail: francesco.marzari@pd.infn.it}
and Gennaro D'Angelo,$^{2}$\thanks{E-mail: gennaro@lanl.gov}
\\
$^{1}$Department of Physics and Astronomy, University of Padova, via Marzolo 8, I-35131, Padova, Italy\\
$^{2}$Theoretical Division, Los Alamos National Laboratory, Los Alamos, NM 87545, USA\\
}

\date{Accepted \today. Received \today; in original form \today}

\pubyear{\the\year}

\begin{document}
\label{firstpage}
\pagerange{\pageref{firstpage}--\pageref{lastpage}}
\maketitle

\begin{abstract}
In evolved and dusty circumstellar discs, two planets with masses 
comparable to Jupiter and Saturn that migrate outwards while maintaining 
an orbital resonance can produce distinctive features in the dust
distribution.
Dust accumulates at the outer edge of the common gas gap, which
behaves as a dust trap, where the local dust concentration is significantly
enhanced by the planets' outward motion. Concurrently, an expanding cavity
forms in the dust distribution inside the planets' orbits, because dust
does not filter through the common gaseous gap and grain depletion in 
the region continues via inward drifting.
There is no cavity in the gas distribution because gas can filter through
the gap, although ongoing gas accretion on the planets can reduce the gas
density in the inner disc. 
Such behaviour was demonstrated by means of simulations neglecting
the effects of dust diffusion due to turbulence and of dust backreaction 
on the gas.
Both effects may alter the formation of the dust peak at the gap outer
edge and of the inner dust cavity, by letting grains filter through
the dust trap.
We performed high resolution hydrodynamical simulations of the coupled 
evolution of gas and dust species, the latter treated as pressureless 
fluids, in the presence of two giant planets. We show that diffusion 
and backreaction can change some morphological aspects of the dust
distribution but do not alter some main features, such as the outer 
peak and the expanding inner cavity.
These findings are confirmed for different parametrizations
of gas viscosity. 
\end{abstract}




\section{Introduction}
\label{sec:Intro}

In a recent study, \cite{marzaridangelo2019} examined 
the distribution of dust particles around two resonant planets
embedded in a circumstellar disc. The two planets, with masses 
comparable to those of Jupiter and Saturn, had orbits in 
a resonant configuration, with ratios of the mean motion
equal to either 2:1 or 3:2.
Because of the type of resonance and of the applied disc
conditions, the planets tend to migrate outwards and
dust particles tend to accumulate outside of the orbit of 
the exterior planet.
Concurrently, the inward migration of dust grains that 
move inside of the orbit of the interior planet leads
to an enlargement of the dust gap compared to the gap in
the gas and to a dynamical decoupling between the gaps in 
the gas and dust distributions.
The build-up of the dust density at the outer
edge of the gap surrounding the planets is markedly higher
in the case of the 2:1 mean-motion resonance and may appear
as a bright ring (at appropriate wavelengths) in resolved
observations of discs.  
A similar phenomenon was also found for lower-mass planets
\citep[in the Super-Earth mass range,][]{marzaridangelo2020},
although less pronounced. All those simulations were performed
without including the effects of possible dust diffusion due
to gas turbulence and of dust back-reaction on the gas.
Here we explore the consequences of these two mechanism on
the accumulation of dust grains at the outer edge of the gap
and on the formation of a wider gap in the dust distribution
compared to the density gap in the gas. This latter feature 
may lead to the formation of a transition disc, if the planets
are close enough to the star (low gas surface density inside
the planets' orbits can be caused by ongoing accretion of 
gas on star and planets). 

Circumstellar discs are likely turbulent \citep[e.g.,][]{hughes2011}.
Various mechanisms have been proposed as drivers of turbulence, 
such as convective instability \citep[e.g.,][]{klahr2014,lyra2014}, 
vertical shear instability \citep[e.g.,][]{urpin2003,nelson2013,stoll2017},
and magneto-rotational instability 
\citep[e.g.,][]{balbus1991,balbus1998,balbus2003}.
Gas turbulence may force dust grains to diffuse (over a length-scale
dictated by the type of turbulence), a process that not only may affect
dust accumulation but can also hinder the efficiency of dust entrapment
at radial location of gas pressure maxima. 
In fact, according to \citet{sierra2019} and \citet{pinilla2020},
dust diffusion may reduce, or even prevent, significant concentrations 
of grains at locations of gas ``bumps'', since the grains can 
disperse out of the region. This process might affect the conclusions
of our previous results on the accumulation of dust at the outer 
edge of the gaseous gap of two planets in resonance by letting
dust seep through the gap and reach the inner disc regions. 
If the effect is large enough, the concentration of dust at 
those radial locations, obtained in the numerical simulations
by \cite{marzaridangelo2019} and \cite{marzaridangelo2020}, may be 
severely depleted or even largely absent.

In addition to dust diffusion, the back-reaction of dust on 
gas can also impact the formation of grain concentration at a local
pressure maximum. According to \cite{taki2016}, dust back-reaction 
can deform the pressure gradient of the gas when high-enough values
of the dust-to-gas mass ratio are reached. This may be the case of 
the dust concentration attained at the outer edge of gaseous gaps,
observed in the simulations of two planets in resonance migrating
away from the star.  

To test the relevance of these two mechanisms, diffusion and
back-reaction, on the formation of dust over-dense regions
caused by the outward migration of two planets in resonance,
we performed simulations of the evolution of two planets
in resonance in which both these two mechanisms are taken 
into account. 
In Section~\ref{sec:MA}, we describe the numerical model exploited
to study the coupled evolution of dust and gas in presence of
the two resonant planets. In Section~\ref{sec:R32}, we outline 
the dust behaviour when the planets evolve in the 3:2 mean-motion
resonance whereas, in Section~\ref{sec:R21}, we analyse the case
of the 2:1 mean-motion resonance. In Section~\ref{sec:AV},
we test the robustness of our results by changing the viscosity
parameterization, including a constant kinematic viscosity 
the one that applies a constant value of the $\alpha$ parameter.
Finally, in Section~\ref{sec:Con}, we discuss our results.

\section{Methods and Algorithms}
\label{sec:MA}

In past work on the coupled evolution of dust and gas in protoplanetary
discs, we adopted a Lagrangian description of the solid component.
Instead, an Eulerian formalism is applied in the present study, since
solids are treated as pressureless fluids. Some details on the involved
equations are provided below to highlight the differences between 
the two approaches.
\citet{marzaridangelo2019} and \citet{marzaridangelo2020} used
the two-dimensional (2D) FARGO hydrodynamics code \citep{masset2000},
modified to include the dynamical evolution of dust particles embedded
in the gaseous disc in  a Lagrangian fashion \citep{picogna2015,picogna2018}.

The drag force on the particles was computed from the local
gas density according to the equation \citep{woitke2003}
\begin{equation}
\mathbf{F} = \left( \frac{3K}{3K+1} \right)^2 \mathbf{F}_{\mathrm{E}} 
           + \left( \frac{1}{3K+1} \right)^2 \mathbf{F}_{\mathrm{S}},
\end{equation}
where $\mathbf{F}_{\mathrm{E}}$ is the Epstein drag contribution, 
given by
\begin{equation}
 \mathbf{F}_{\mathrm{E}} = - \frac{4}{3} \left( 1 + \frac{9 \pi }{128} M^2 \right)^{1/2} s^2 \rho_{g} v_{\mathrm{th}} \mathbf{v}_{\mathrm{rel}},
\end{equation}
and $\mathbf{F}_{\mathrm{S}}$ is the Stokes drag component,
given by
\begin{equation}
 \mathbf{F}_{\mathrm{S}} = - \frac{1}{2} C_D  \pi s^2  \rho_{g} v_{\mathrm{rel}} \mathbf{v}_{\mathrm{rel}}.
\end{equation}
In the above equations, $\rho_{g}$ is the gas density, $s$ is 
the radius of the particle, $v_{\mathrm{th}}$ is the local thermal
velocity of the gas and $\mathbf{v}_{\mathrm{rel}}$ is the velocity
of the dust particle relative to the gas. The quantity $K$ is 
the Knudsen number and $M$ is the Mach number (computed from 
the particle's relative velocity $v_{\mathrm{rel}}$).
Quantity $C_D$ is the drag coefficient for the Stokes
regime.

In this paper, to test the relevance of diffusion and
back-reaction on the formation of dust over-dense regions
caused by the outward migration of two planets in resonance,
we carry out simulations with the code FARGO3D \citep{llambay2019}.
In this version of the code, the dust particles are treated
as additional pressureless fluids where momentum is transferred
between the gas and each of the dust species (but not among
dust species). 
The dust fluid is affected by Epstein drag, which imparts a force 
\emph{per unit volume} to a dust species given by
\begin{equation}
 \mathbf{F}_d =  - \rho_{d}\frac{\Omega}{\tau_s} (\mathbf{v}_{d} - \mathbf{v}_{g}),
 \label{eq:Fd}
\end{equation}
where $\rho_{d}$ is the dust density, $\tau_s$ is the Stokes number 
of the particle and $\Omega$ is the Keplerian frequency of the gas. 
An equal and opposite force is imparted to the gas.
Note that, in Equation~(\ref{eq:Fd}), information regarding
the drag coefficient are incorporated into $\tau_s$.

A term is added to the continuity equation to model the diffusion
of the dust species within the gas
\citep{morfill1984}
\begin{equation}
\frac {\partial \rho_{d}} {\partial t} = \nabla \cdot \left ( D_{d} \rho \nabla \frac {\rho_{d}} {\rho} \right ),
\label{eq:diff}
\end{equation}
where $\rho = \rho_{g} + \rho_{d}$ and $\rho_{d}$ is the density of
individual dust species.
Equation~(\ref{eq:diff}) is only applied to the pressureless fluids and
it assumes the same diffusion coefficient for all dust species, which is
taken equal to the value of the gas kinematic viscosity
\citep{brauer2008}
\begin{equation}
D_{d} = \nu.
\label{eq:Dd}
\end{equation}
The effects of this choice are not tested.

The original version of the code applies Stokes numbers,
$\tau_{s}$, in Equation~(\ref{eq:Fd}) that are constants.
The code was modified so that we can select each pressureless fluid
(dust species) not according to a Stokes number, which varies as
a function of the local properties of the gas (density, temperature 
and velocity), but rather according to the particle size. 




\section{Dust distribution near planets in the 3:2 resonance}
\label{sec:R32}

We first investigate the case of a pair of planets that become
captured in the 3:2 mean-motion resonance and migrate outward
thereafter.
The interior planet has the mass of Jupiter whereas the exterior
planet has the mass of Saturn.
A more massive inner planet is a condition conducive to outward
migration.
The planets orbit in a cold, local-isothermal disc with a fixed
aspect ratio $H/r=h=0.02$. 
(Calculations with a larger ratio, $h=0.05$, are also presented.)

The disc extends from $0.4$ to $12\,\mathrm{au}$ and is discretised
over a grid of $512\times 1024$ area elements (in the radius and
azimuth, respectively). The initial surface density of the gas 
declines as 
\begin{equation}
\Sigma(r) = \Sigma_{0}\left(\frac{r_{0}}{r}\right),
\end{equation}
where $\Sigma_{0} = 200\,\mathrm{g\,cm^{-2}}$ is the density at 
the reference radius $r_{0}=1\,\mathrm{au}$.

Three different populations of icy grains (bulk density
$1\,\mathrm{g\,cm^{-3}}$) are included in the simulations,
whose sizes are $100\,\mu\mathrm{m}$, $1\,\mathrm{mm}$ and $1\,\mathrm{cm}$.
For the applied disc conditions, these particles have Stokes numbers
less than $\approx 0.1$.

The initial dust-to-gas mass ratio for each of the three dust species
is $0.0033$, so that the overall dust-to-gas mass ratio adds up to 
$0.01$, which is a typical value adopted for circumstellar discs
and is based on values found in the interstellar medium. 
However, dust needs not be primordial in origin, that is,
part of the inventory of solids from which the planets formed. 
In fact, the dust populations may represent, or contain, second 
generation dust produced by collisions among leftover planetesimals, 
after the planets became massive enough
\citep{turrini2019,gennaro2022,bernabo2022}. 
The equations describing the evolution of the system are solved 
in a non-inertial reference frame centered on the star, including
the indirect terms arising from the planets' and disc's gravity.

\begin{figure}
\includegraphics[width=0.7\columnwidth, angle=-90]{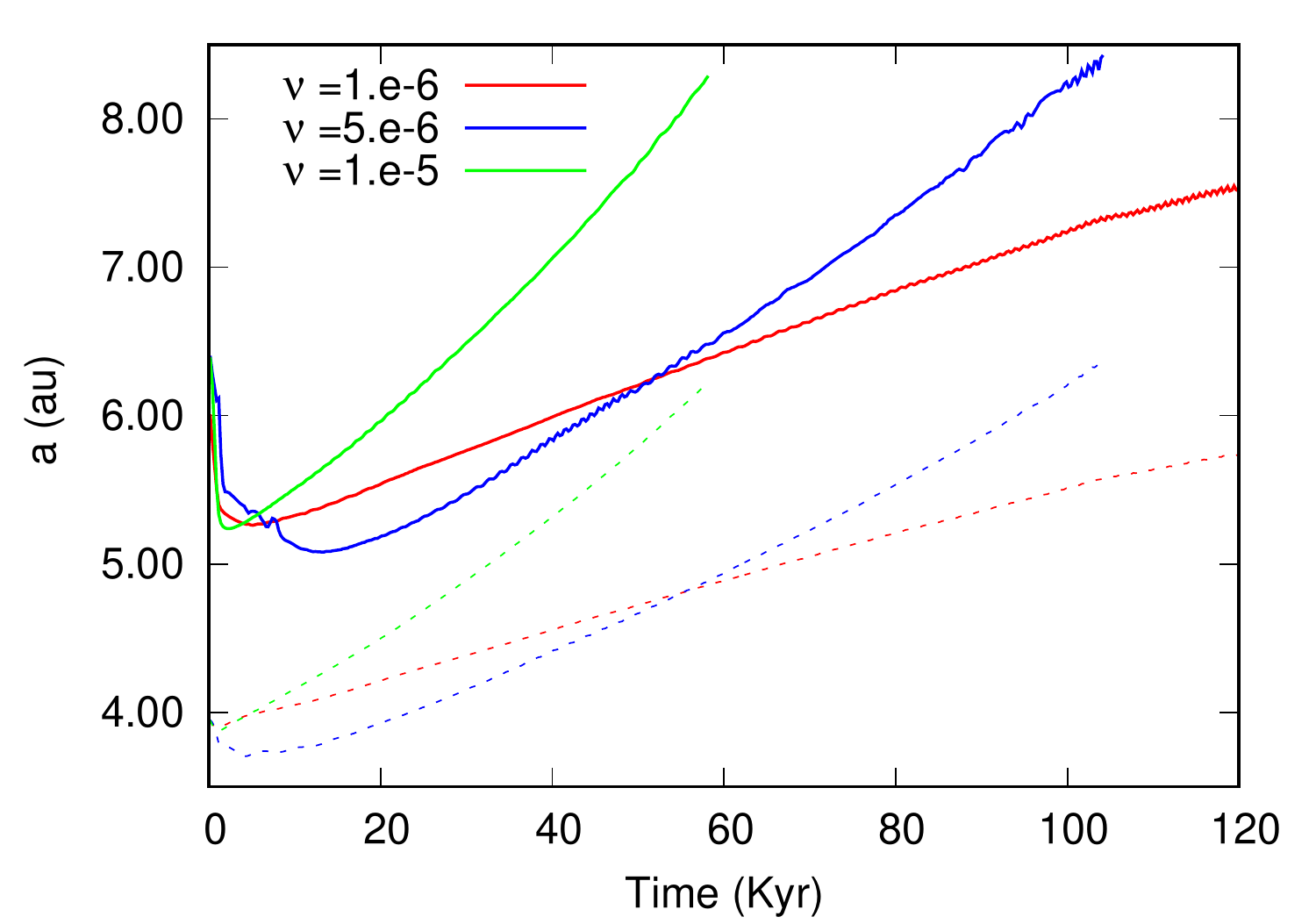}
\includegraphics[width=0.7\columnwidth,angle=-90]{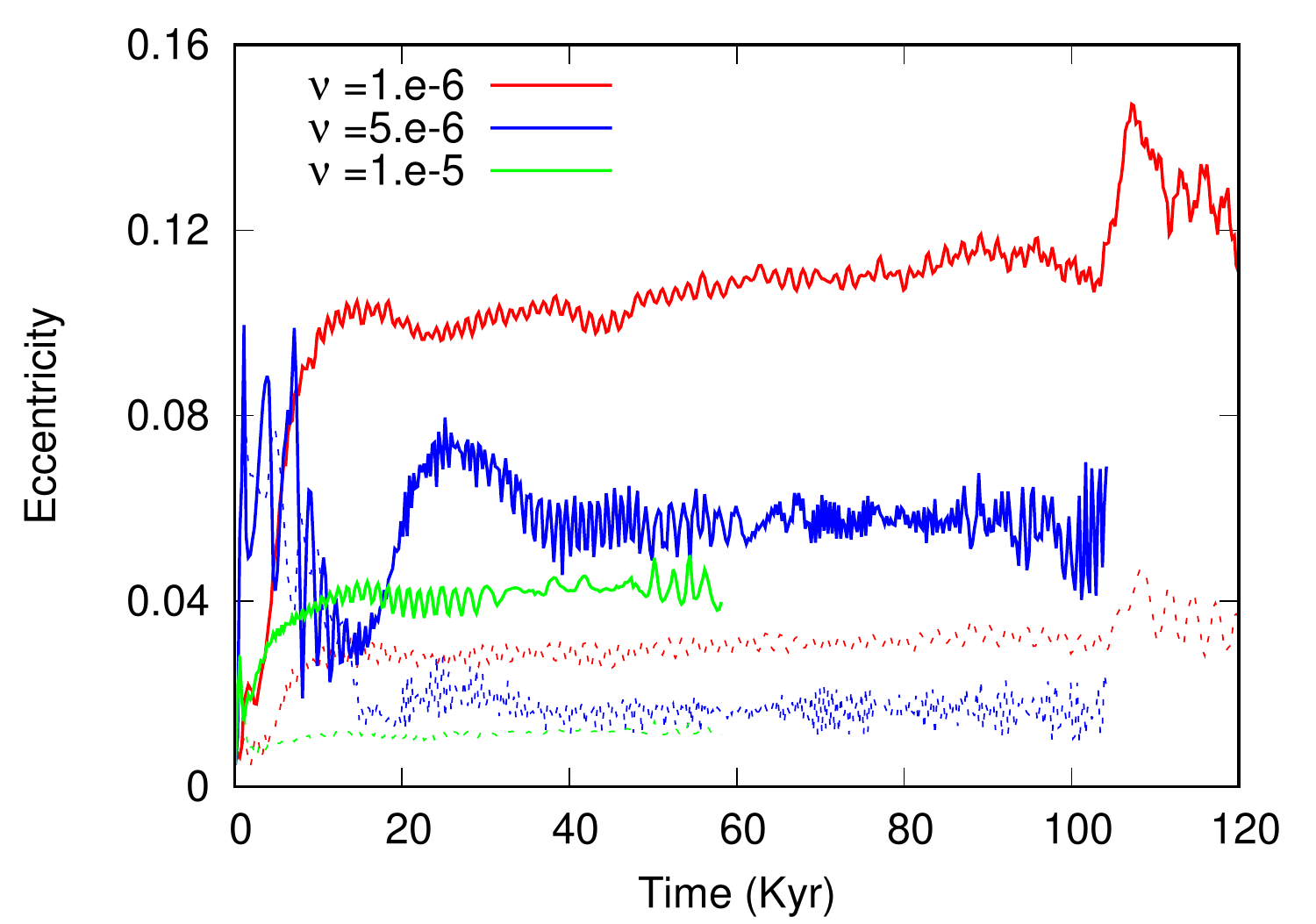}
\caption{The top panel shows the semi-major axis of pairs of
planets during their migration, for three values of a constant
kinematic viscosity, $\nu$, of the gas 
(in units of $r^{2}_{0}\Omega_{0}$, see text).
In these cases,
the planet pair is locked in the 3:2 mean-motion resonance.
In the bottom panel illustrates the evolution of the orbital eccentricity.
The inner, more massive planet has a lower eccentricity in
all cases. 
Data are averaged over a $250$~yr window; see text for further details.
}
\label{fig:mig_speed_32}
\end{figure}

Various values of the gas kinematic viscosity, $\nu$, are
considered because this parameter affects the tidal interactions
between the planets and the gas, and also determines the amount
of dust diffusion through Equation~(\ref{eq:Dd}).
Additionally, it can also affect the speed of the planets' outward
migration.
In these models, we adopt a constant value of kinematic viscosity.
The impact of $\nu$ on the efficiency of the outward migration 
is illustrated in Figure~\ref{fig:mig_speed_32}, for a given value
of the initial gas density at the reference radius $r_{0}$.
In these models, the planets start to migrate at the beginning
of the simulations, when the distributions of gas and dust are
unperturbed (hence the initial rapid inward migration of the
planets).
As the outer planet approaches the inner planet and their 
orbits become caught in resonance, the tidal perturbations
exerted by the outer planet on the circumstellar material
alter the torque balance on the inner planet. Consequently,
the inner planet first slows down and then migrates away 
from the star, pushing the outer planet outward through
the resonance forcing.

The evolution of the semi-major axis of the outer planet, $a_{2}$,
is shown for three different values of the gas kinematic viscosity: 
$\nu = 10^{-6}$, $5 \times 10^{-6}$, and $10^{-5},$
in units of $r^{2}_{0}\Omega_{0}$ ($\Omega_{0}$ is the Keplerian
frequency at $r_{0}$).
Note that a constant kinematic viscosity corresponds to a variable
$\alpha$ parameter, $\alpha\propto \nu/(h^{2}\sqrt{r})$.
With our choice of parameters, at $r=r_{0}$, said values of $\nu$
would correspond to $\alpha_{0}=0.0025$, $0.0125$, and $0.025$,
respectively.
However, in the disc regions where the planets orbit, $\alpha$ would
be smaller by a factor of $2$ or more.

After a different behaviour at the beginning of the evolution,
prior to or shortly after the capture into the 3:2 orbital resonance
(see Figure~\ref{fig:mig_speed_32}),
the planets undergo sustained outward migration locked in mean-motion
resonance. The migration speed of the pair is related to $\nu$ and
determined by the shape of the common (or overlapping) gaseous gap
of the two planets.
A similar outcome is obtained for the cases involving the 2:1
resonance, as shown in the next section. 
Notice that the outer planet is subject to a negative torque
exerted by the disc material exterior to its orbit, hence 
it would tend to move inward, whereas the resonance forcing 
pushes it outward.
These two opposing torques allow the resonance to be maintained
during the outward migration of the pair.

In the calculations presented herein, all disc 
material exerts torques on the planets, including material within
the planets' Hill spheres. Since the numerical resolution
is limited and density variations close to the planets may not 
be properly resolved, some spurious effects may arise that affect
outward migration.
To quantify possible differences, one model was also performed
by removing the torques exerted from within the planets'
Hill spheres. The orbital evolution is comparable to that of
the calculation with default setup, although at a somewhat 
reduced outward migration speed.
The amount of kinematic viscosity can also alter the local
distribution of material around the planets and unresolved density
gradients can possibly impact the resulting migration velocity.
Nonetheless, it must be pointed out that for the purposes 
of this study the details of the outward migration process 
are not important, henceforth tests on the response of the system
to numerical parameters are unnecessary.
The only requirement is that the planet pair becomes locked 
in resonance and moves away from the star for a prolonged amount 
of time.

\begin{figure}
\includegraphics[width=0.75\columnwidth, angle=-90]{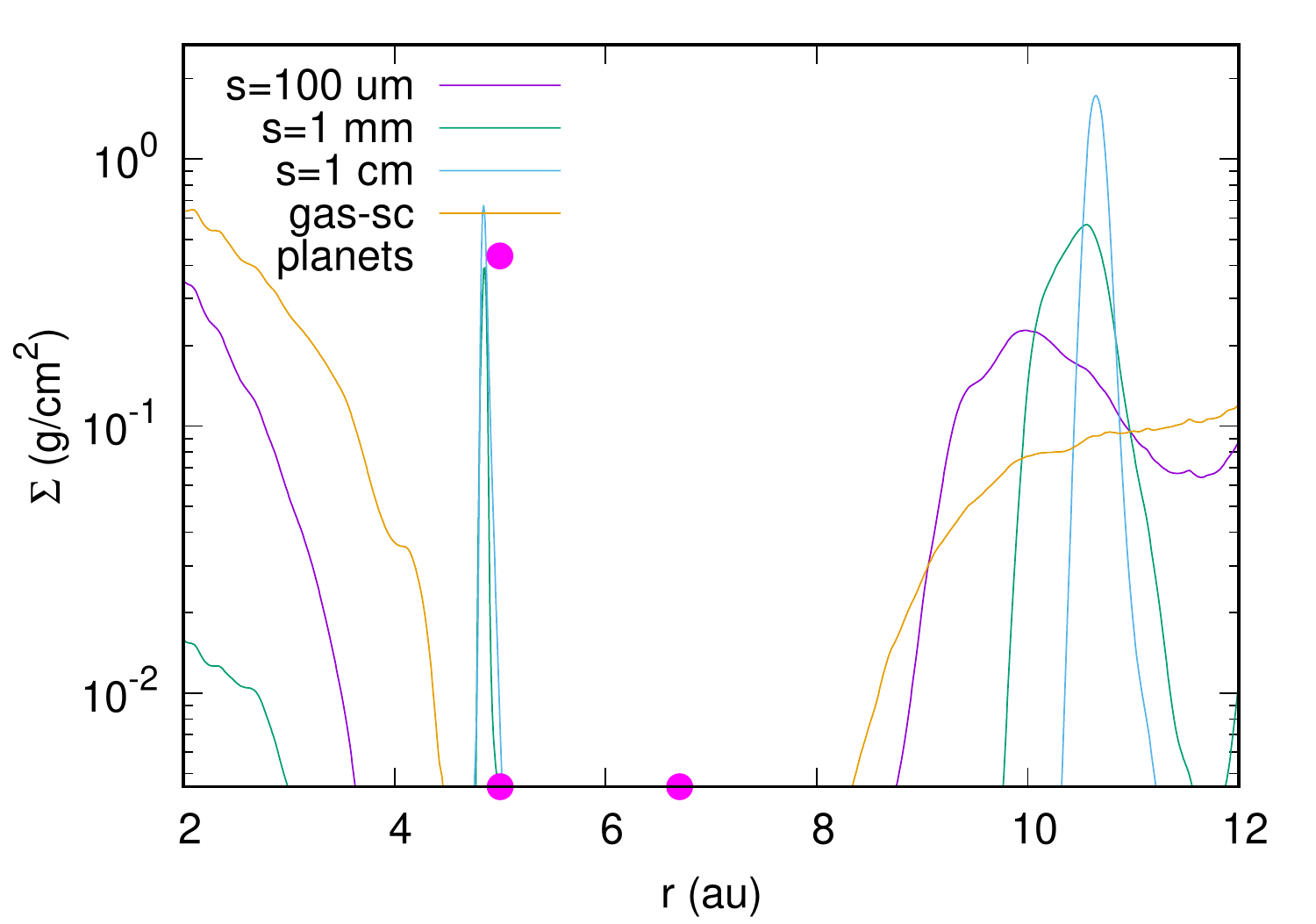}
\includegraphics[width=0.75\columnwidth,angle=-90]{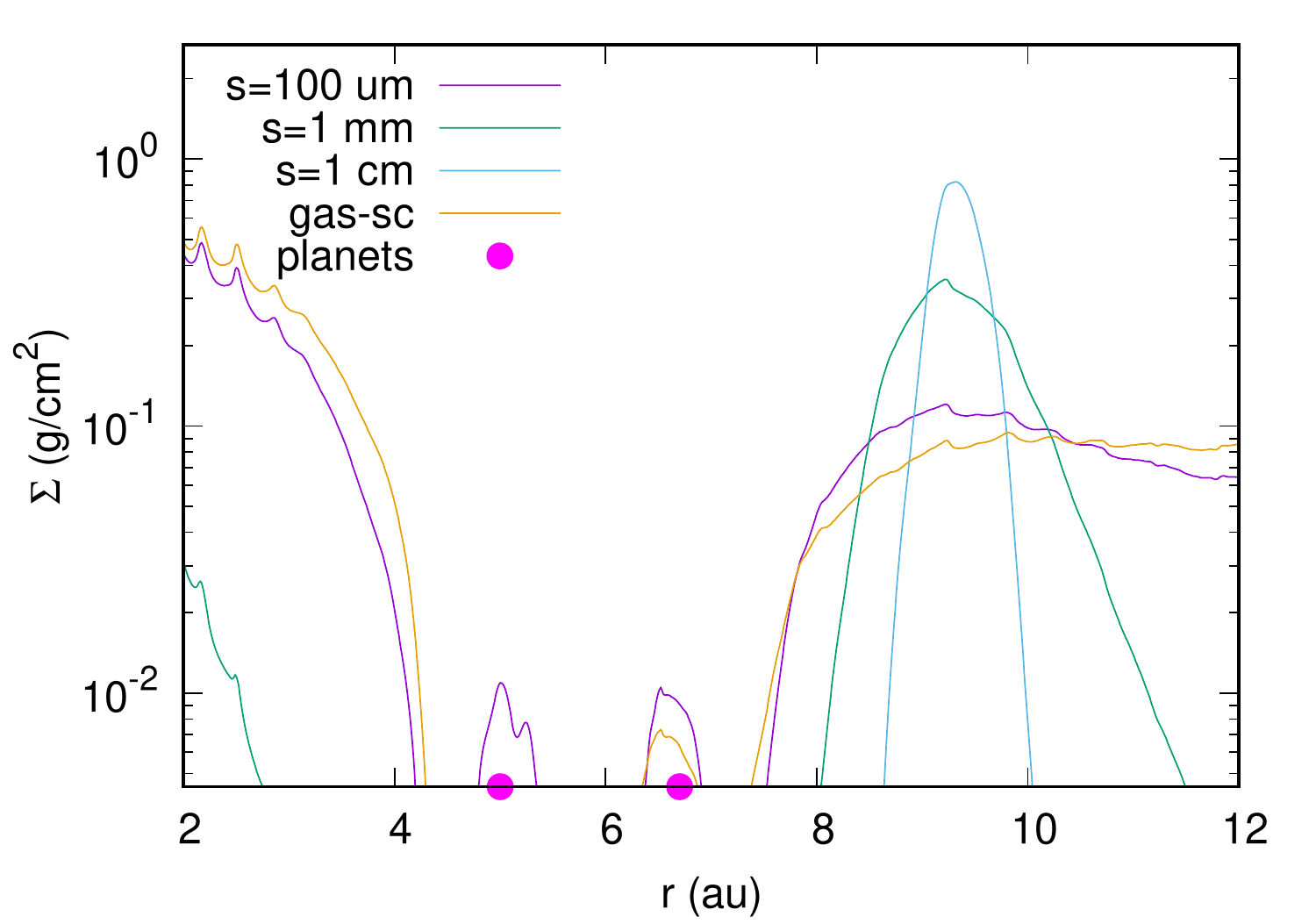}
\includegraphics[width=0.75\columnwidth,angle=-90]{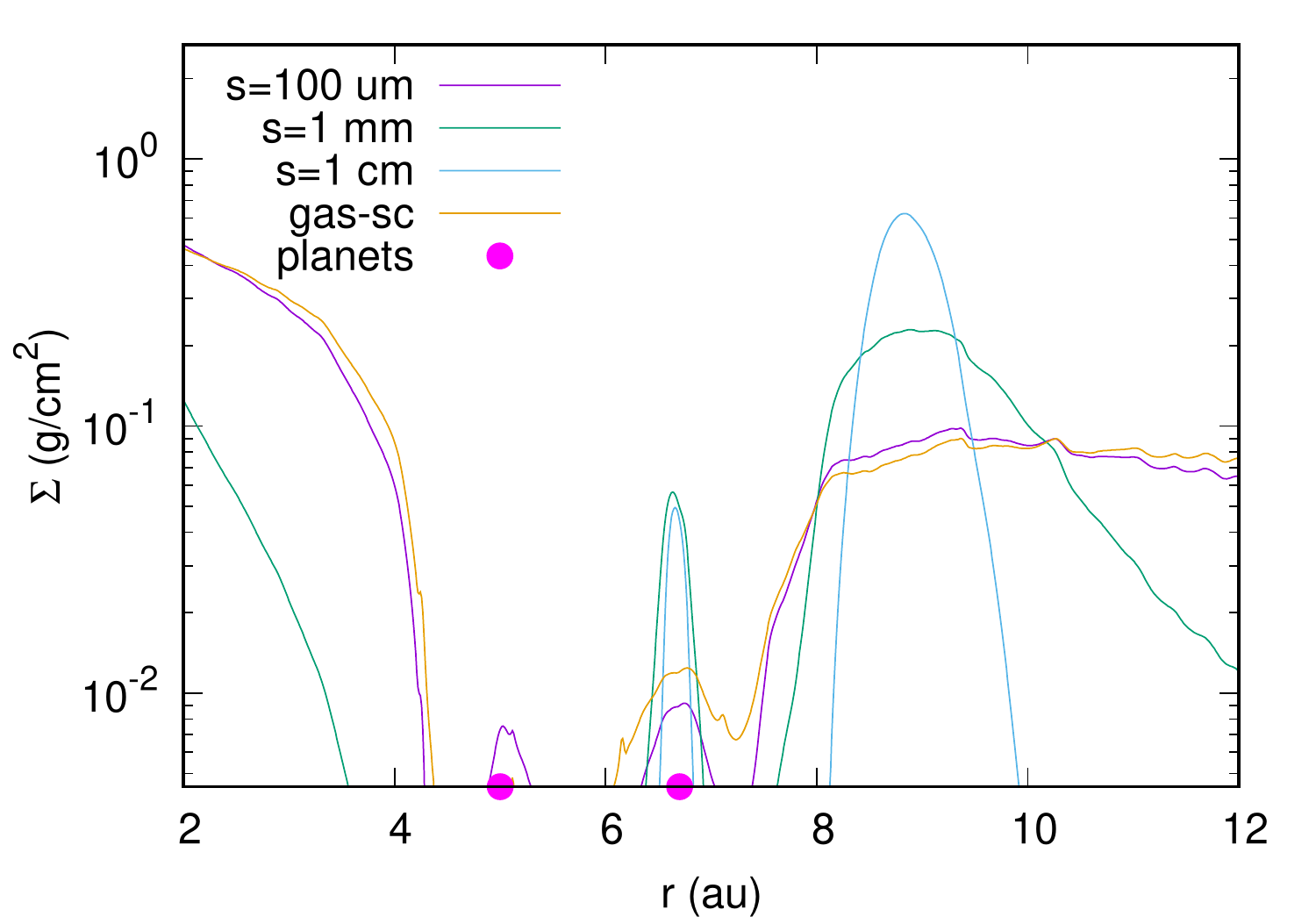}
\caption{Surface density (averaged in azimuth around the star)
of dust particles of different sizes, ranging from 100$~\mu$m to $1$~cm.
These profiles are compared to the re-scaled gas density 
(i.e., multiplied by $0.0033$; the total dust-to-gas mass ratio
is $0.01$).
The top panel refers to a kinematic viscosity equal to 
$\nu = 10^{-6}$, the middle panel to
$\nu = 5 \times 10^{-6}$ and the bottom panel to
$\nu = 10^{-5}$, in units of $r^{2}_{0}\Omega_{0}$.
}
\label{fig:prof_nu_32}
\end{figure}

When the outer planet reaches $6.5$~au from the star, we compare
the dust distributions in the three cases with different viscosity.
This comparison is shown in Figure~\ref{fig:prof_nu_32}. 
The dust density profiles show a peak at the outer border of 
the gas gap carved by the planets' tidal perturbations.
This peak is more marked in the distributions of the largest
grains, which are less coupled to the gas (i.e., they have
a larger Stokes number and therefore a longer coupling timescale).
Dust-to-gas mass ratios at the peak location range from $0.04$ to
$0.08$, increasing as $\nu$ decreases.
In the region inside the inner edge of the gap there is a significant
depletion of dust due to drifting motion towards the star.
Re-supply of dust to this region is reduced, or halted, by 
the dust trap at the outer edge of the gas gap. After some time,
the disc would appear as a transition disc with an inner cavity 
in the dust density, which expands outward over time due to 
the outward migration of the planets.
These effects appear more evident at lower gas viscosity, which may
be due to the lower level of diffusion but it may also be related
to the different migration velocity of the planet pair 
(see Figure~\ref{fig:mig_speed_32}), as discussed below.
Beyond $\approx 10$~au the disc is depleted of mm- and cm-grains
(but not of the smallest grains), an effect associated to the inward 
drift of the particles via gas drag (which does not affect as much 
the smallest grains). This is a boundary effect related to the fact
that particles are not flowing inward from greater distances
(i.e., there is not re-supply of solids at the outer boundary).
Test simulations not reported here confirm this issue.

The accumulation efficiency of particles at the outer edge of 
the gas gap depends on the stopping time of the particles 
($\tau_s/\Omega$) and
the timescale over which the radial pressure gradient of the gas
moves (in our case, due to planet migration). For a given
stopping time, the shorter the outward migration timescale,
the less time dust grains have to accumulate. Stated differently,
for a similar orbital configuration of the planets, i.e., similar
orbital frequencies $\Omega$, a more rapid outward migration 
can facilitate the filtering process of dust toward
the star. The overall outcome is a less depleted inner disc
in the cases with more vigorous outward migration.

For the same reason, the different migration speed also controls
the build up of the dust at the outer edge of the gas gap. 
As a consequence, the reduction in the peak density at the outer 
edge of the gap observed at higher viscosity values, shown in 
Figure~\ref{fig:prof_nu_32}, is likely due to the combination of
the two effects: a higher diffusion rate and a decrease in the
trapping efficiency of dust grains caused by the faster outward
migration rate of the planets. 

\begin{figure}
\includegraphics[width=0.75\columnwidth, angle=-90]{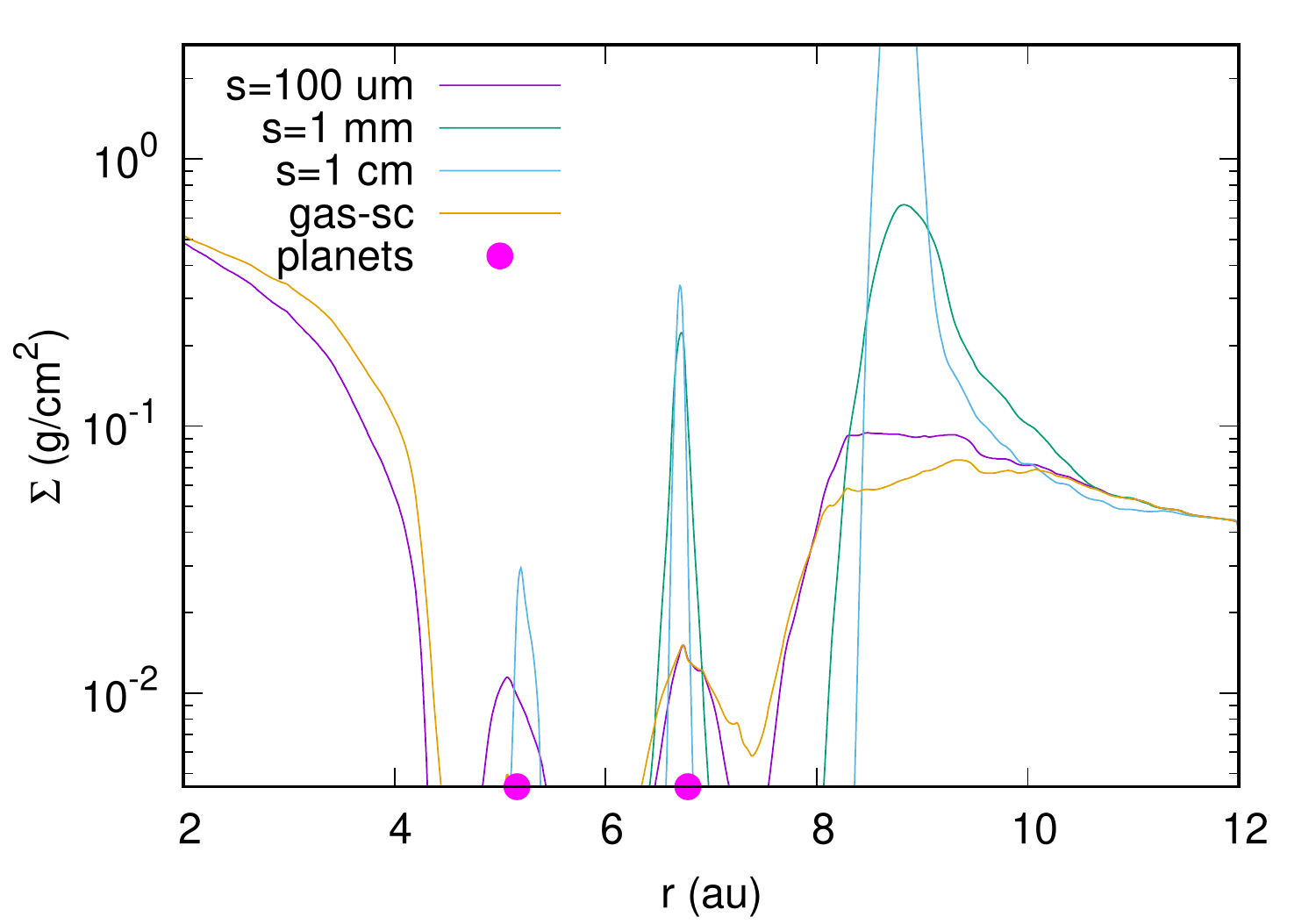}
\caption{Dust and gas profiles (as in 
  Figure~\ref{fig:prof_nu_32}) with  $\nu = 10^{-5}$~$r^{2}_{0}\Omega_{0}$,
  but without the inclusion of diffusion and back-reaction.}
\label{fig:prof32_nodiff}
\end{figure}

To better characterize the role of diffusion and back-reaction
in the formation, size and shape of the outer peak in dust density, 
we ran a simulation in which both these two processes were
neglected, adopting a kinematic viscosity 
$\nu = 10^{-5}$~$r^{2}_{0}\Omega_{0}$ (see Figure~\ref{fig:prof32_nodiff}).
In this model, the migration speed is broadly consistent with
that of the model illustrated in the bottom panel of 
Figure~\ref{fig:prof_nu_32}, but the peak in the dust density
distributions of the largest particles tend to be higher and
sharper compared to corresponding one in the the bottom panel
of Figure~\ref{fig:prof_nu_32}. 
The implication is that diffusion is indeed affecting 
the concentration of the dust by spreading large grains over 
a broader radial region. 
Nonetheless, the reduced efficiency in collecting dust caused by 
diffusion is not strong enough to prevent the formation of 
a prominent density peak outside the planets' orbits,
which may be detected by high resolution observations. 

\begin{figure}
\includegraphics[width=\columnwidth,trim={0mm 8mm 0mm 0mm},clip]{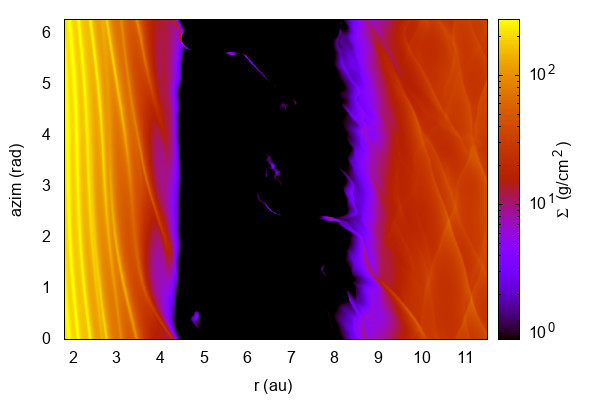}
\includegraphics[width=\columnwidth,trim={0mm 8mm 0mm 0mm},clip]{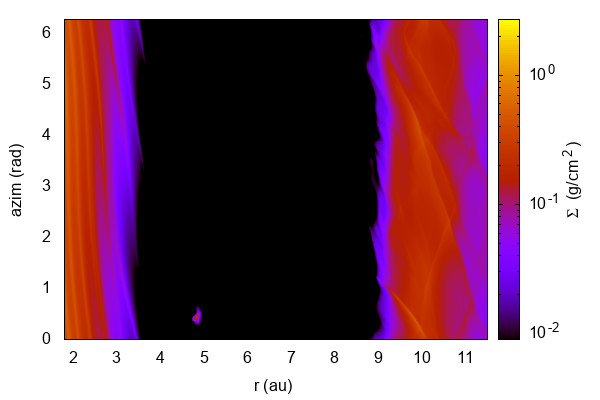}
\includegraphics[width=\columnwidth,trim={0mm 8mm 0mm 0mm},clip]{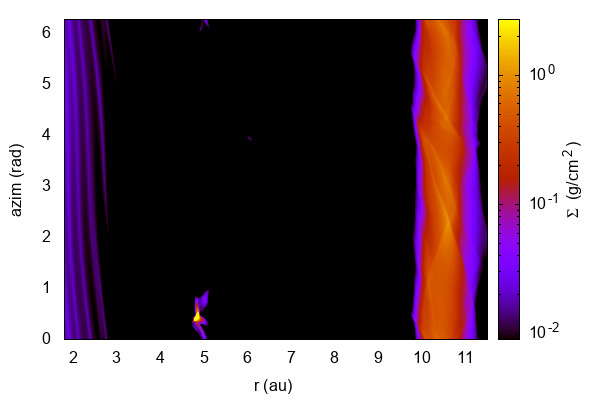}
\includegraphics[width=\columnwidth,trim={0mm 0mm 0mm 0mm},clip]{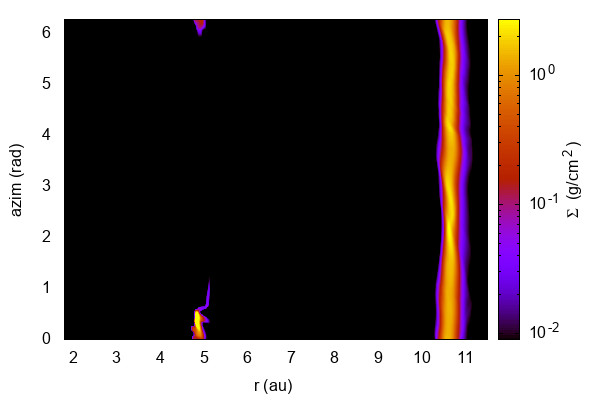}
\caption{Surface density maps illustrating gas and dust
distributions around two planets locked in the 3:2 resonance
($\nu = 10^{-6}$~$r^{2}_{0}\Omega_{0}$).
The top panel shows the gas density, whereas the second, third 
and fourth panels represent $100$~$\mu$m, $1$~mm and $1$~cm particles,
respectively. 
        }
\label{fig:map32}
\end{figure}

In Figure~\ref{fig:map32} we plot the surface density distribution 
of the gas and of dust grains of different sizes for the case with
lowest viscosity, $\nu = 10^{-6}$~$r^{2}_{0}\Omega_{0}$.
The gas gap is significantly narrower than that of the dust
and the width of the latter is larger for larger grain size.
For $1$~cm size dust grains, the density distribution is confined
in an over-dense ring at the outer edge of the gas gap.
Therefore, according to these results, dust diffusion and 
back-reaction are not able to prevent the formation of narrow rings
in the dust distribution at the outer edge of the gas gap. 
In our simulations there is no re-supply of dust at the outer 
boundary of the simulated disc, but it is expected that a continuous 
distribution of solids beyond the grid boundary would supply dust
to the inner disc regions. In this case, at the outer edge
of the gap, we would observe an enhanced density, as predicted
by our simulations. Beyond this density peak, however, there 
would be a continuous distribution of dust originating from 
more distant regions.
Over timescales much longer than those simulated here, dust 
drifting from larger distances may also affect the density peaks
at the outer edge of the gas gap.
In fact, an additional simulation with a wider radial boundary
(not reported here) does show enhanced peaks in large grains,
due to solids drifting from farther distances. Also the population
of small grains would increase at the peaks over longer times,
but at a slower rate, dictated by the drift velocity.
Nonetheless, the depletion of dust within the inner edge of 
the gas gap would not be affected by this process because 
the outer dust trap appears to be efficient enough to halt
(or severely impede) refilling of grains. 
In fact, if refilling of the inner disc was sustained, 
it would occur within the time of our simulations but 
it is not observed.

Continued supply of dust from farther out in the disc may,
at some point, raise the density in the peak regions beyond
some threshold value to make the peaks unstable. For example,
the dust-to-gas mass ratio may become large enough to induce
a back-reaction response on the gas that redistributes the particles
over some radial region via collisions and/or enhanced dust 
coagulation may ensue
(coagulation into larger particles would reduce the back-reaction
force exerted on the gas, because of the lower surface-to-mass ratio).
The back-reaction response may also smooth out the gas density
gradient in the radial direction, altering the radial pressure
gradient and reducing its ability to retain grains.
Such processes are not considered in this study.

\begin{figure}
\includegraphics[width=0.75\columnwidth, angle=-90]{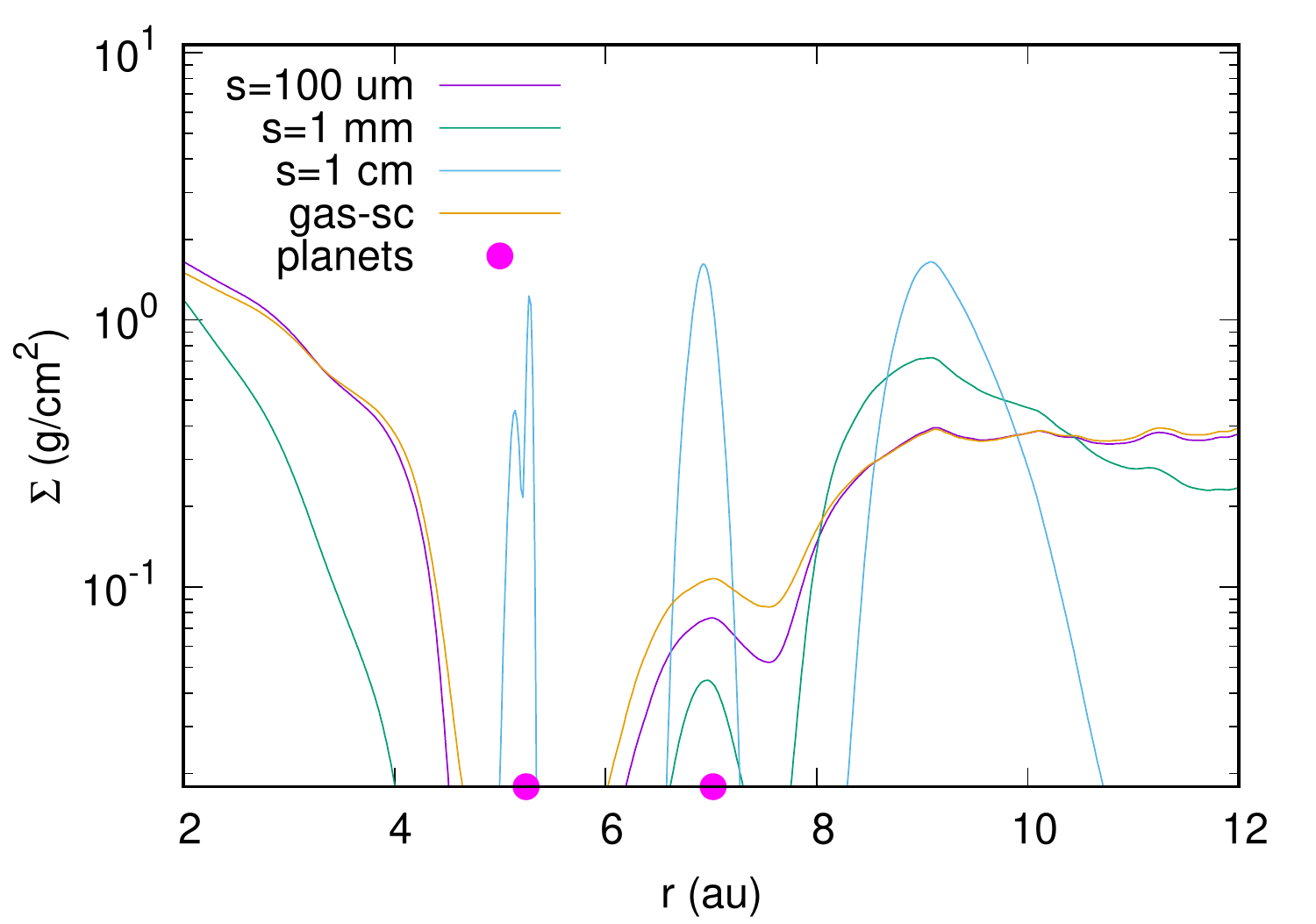}
\includegraphics[width=0.75\columnwidth,angle=-90]{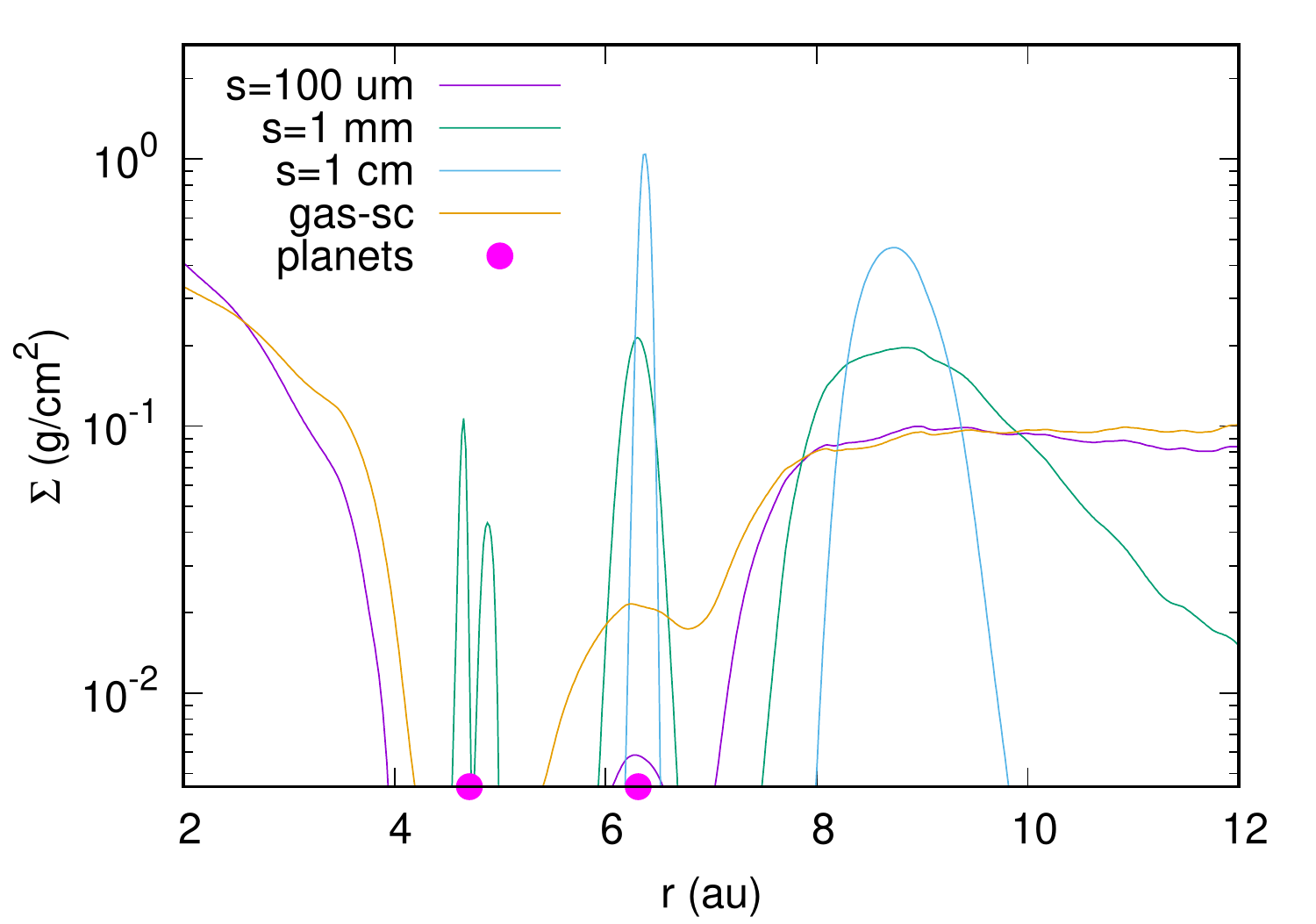}
\caption{   
         Dust and gas density profiles (as in Figure~\ref{fig:prof_nu_32})
         in a disc with $\nu = 10^{-5}$~$r^{2}_{0}\Omega_{0}$ and aspect ratio
         $H/r=h=0.05$. In the top panel $\Sigma_{0} = 800\,\mathrm{g\,cm^{-2}}$
         whereas $\Sigma_{0} = 200\,\mathrm{g\,cm^{-2}}$ in the bottom panel.}
\label{fig:prof_nu_32_H05}
\end{figure}

To test the impact of a larger aspect ratio (i.e., warmer disc)
on the formation of the inner dust cavity and of the peak external to 
the gas gap, we performed two additional simulations adopting $h=0.05$.
In the first model we used a higher gas surface density 
($\Sigma_{0} = 800\,\mathrm{g\,cm^{-2}}$), in order to increase the speed of planet,
migration while in the second we used the same density as in the previous
cases ($\Sigma_{0} = 200\,\mathrm{g\,cm^{-2}}$).
In the latter simulation, since the rate of outward migration is smaller,
the planets are located closer to the star in the plot.
In the top panel of Figure~\ref{fig:prof_nu_32_H05}, the density profiles
are shown for the different grain sizes in the high density case. 
The bottom panel illustrates those profiles for the low density case.
In both simulations the density patterns are similar to those in 
Figure~\ref{fig:prof_nu_32}, suggesting that a higher aspect ratio does
not impair the ability of two planets in resonance to carve an inner gap
in the distribution of the larger grains and to trap dust at 
the gap's outer edge (dust-to-gas mass ratios around the peak region
are $0.02$--$0.03$). 
Thus, dust filtering through the outer edge of 
the gas gap is not increased by the different morphology of the dust 
trap in these warmer discs.

\section{Dust distribution near planets in the 2:1 resonance}
\label{sec:R21}

To test the evolution of the dust when the planets are captured
in a 2:1 mean-motion resonance, we decreased the gas density to
$\Sigma_{0} = 40\, \mathrm{g/cm^2}$ in order to induce orbital
locking in this resonance. It is known that capture in this
mean-motion resonance is a more delicate process than capture
in the 3:2 resonance. If the relative migration velocity with
which the pair of planets approach each other is above a certain
threshold, the resonance forcing is overcome and convergent
migration continues. Additionally, once the 2:1 orbital resonance
is established, migration of the two planets typically proceeds
inward because of the way the two gas gaps overlap. In this
configuration, outward migration may be obtained by choosing
an appropriately low kinematic viscosity so that a wide, 
common gaseous gap forms around the orbits of the two planets.

\begin{figure}
\includegraphics[width=0.7\columnwidth,angle=-90]{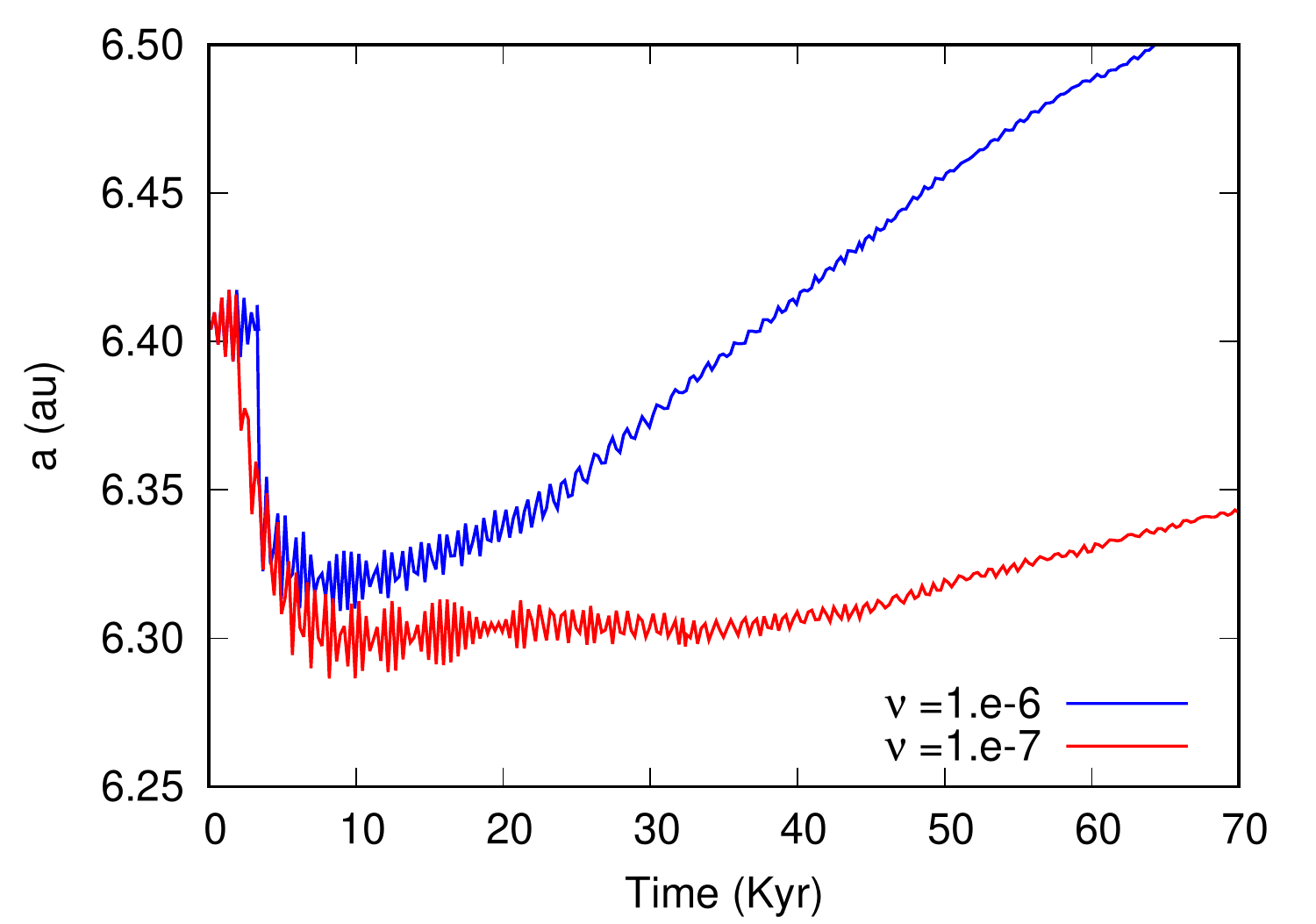}
\includegraphics[width=0.7\columnwidth,angle=-90]{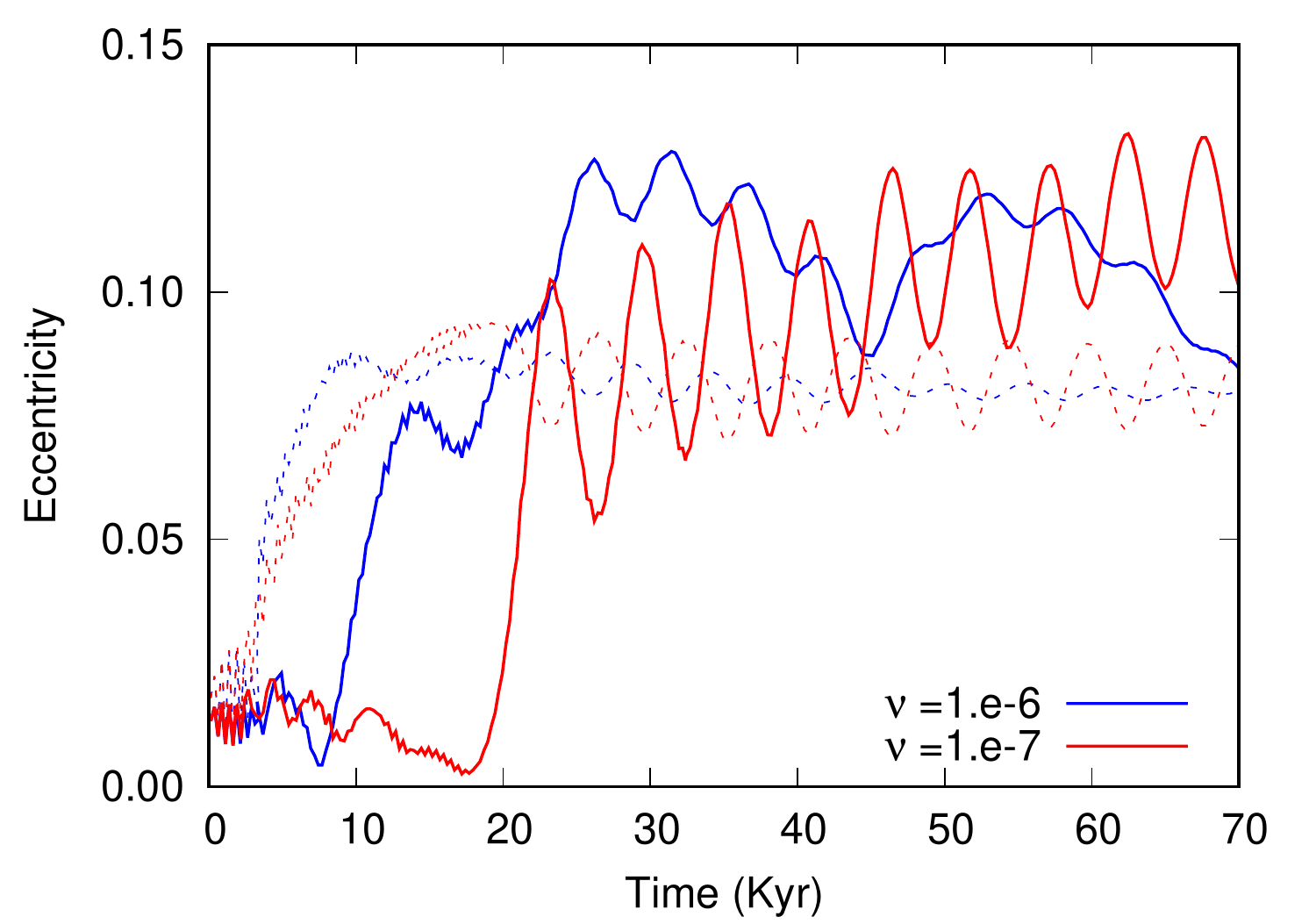}
\caption{
The top panel shows the migration of the exterior
planet of a pair, for two values of a constant kinematic viscosity
of the gas, $\nu$, as indicated in the legend.
The planet pair is locked in the 2:1 mean-motion resonance.
Only the semi-major axis of the exterior planet is plotted because
of the larger mutual distance compared, for example, to the 3:2 case.
The orbital eccentricity of both interior and exterior planets
is illustrated in the bottom panel.
Data are averaged over a $250$~yr window as in the case of the 
3:2 resonance (Figure~\ref{fig:mig_speed_32}).
}
\label{fig:mig_speed_21}
\end{figure}

As in the the models of the previous section, the planets start 
to migrate in unperturbed distributions of gas and therefore 
the initial inward migration of the planets is artificially rapid.
This choice affects the radius at which the planets are
trapped in resonance but is otherwise not much relevant for
the purposes of this study.
As for the models discussed in Section~\ref{sec:R32}, the model
with a larger value of the kinematic viscosity results in a more
vigorous outward migration of the planets once they become trapped
in resonance.
This is illustrated in Figure~\ref{fig:mig_speed_21} for cases with
$\nu = 10^{-6}$ and $\nu = 10^{-7}$~$r^{2}_{0}\Omega_{0}$
($\alpha_{0}=2.5\times 10^{-3}$ and $2.5\times 10^{-4}$, respectively)
.
Note, however, that the orbits also become significantly eccentric,
which also affects outward migration \citep[][]{gennaro2006}.
We also tested a larger value, $\nu = 5 \times 10^{-6}$~$r^{2}_{0}\Omega_{0}$,
but the outer planet crosses the 2:1 resonance with the inner planet.
The pair is temporarily trapped in the 5:3 resonance at which 
point it begins migrating outwards.
That resonance is then broken and the pair becomes finally captured
in the 3:2 resonance, continuing to migrate outward.
For even higher values of the kinematic viscosity, there is no trapping
in the 2:1 resonance, which might be attained by further reducing
$\Sigma_{0}$. However, this possibility was not tested since it
could lead to initial dust-to-gas mass ratios too dissimilar from
the other simulations presented herein.

\begin{figure}
\includegraphics[width=0.75\columnwidth, angle=-90]{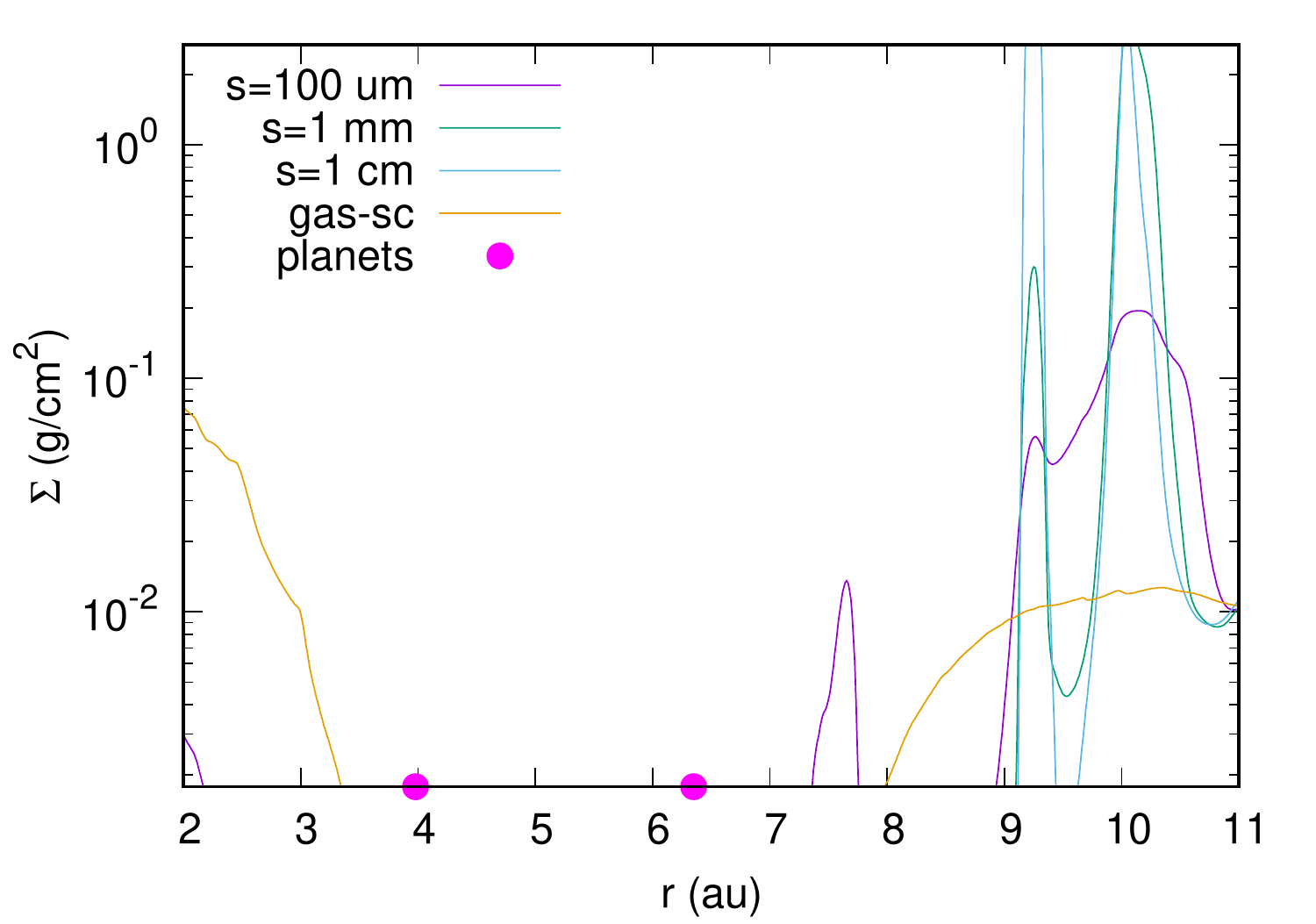}
\includegraphics[width=0.75\columnwidth,angle=-90]{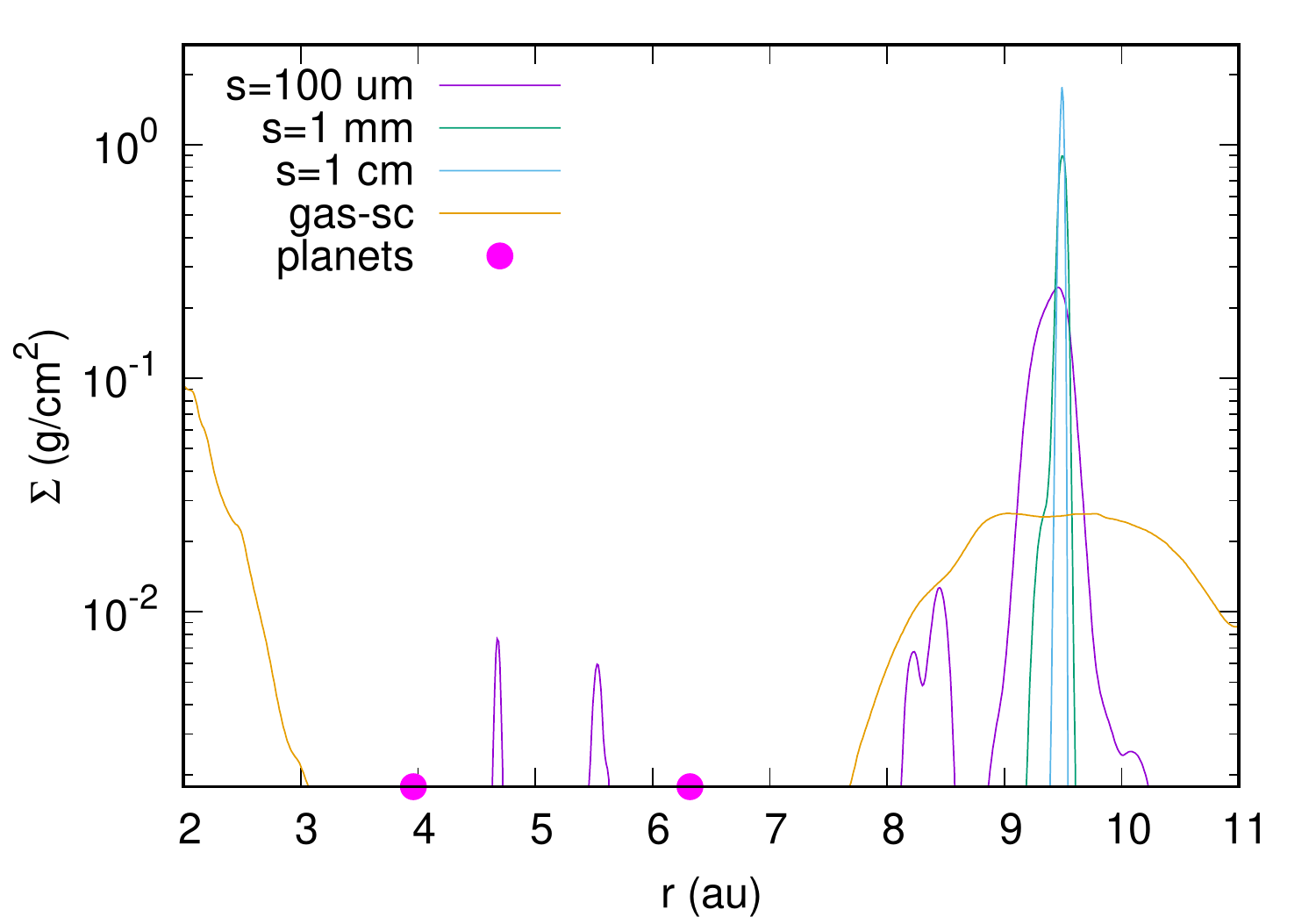}
\includegraphics[width=0.75\columnwidth,angle=-90]{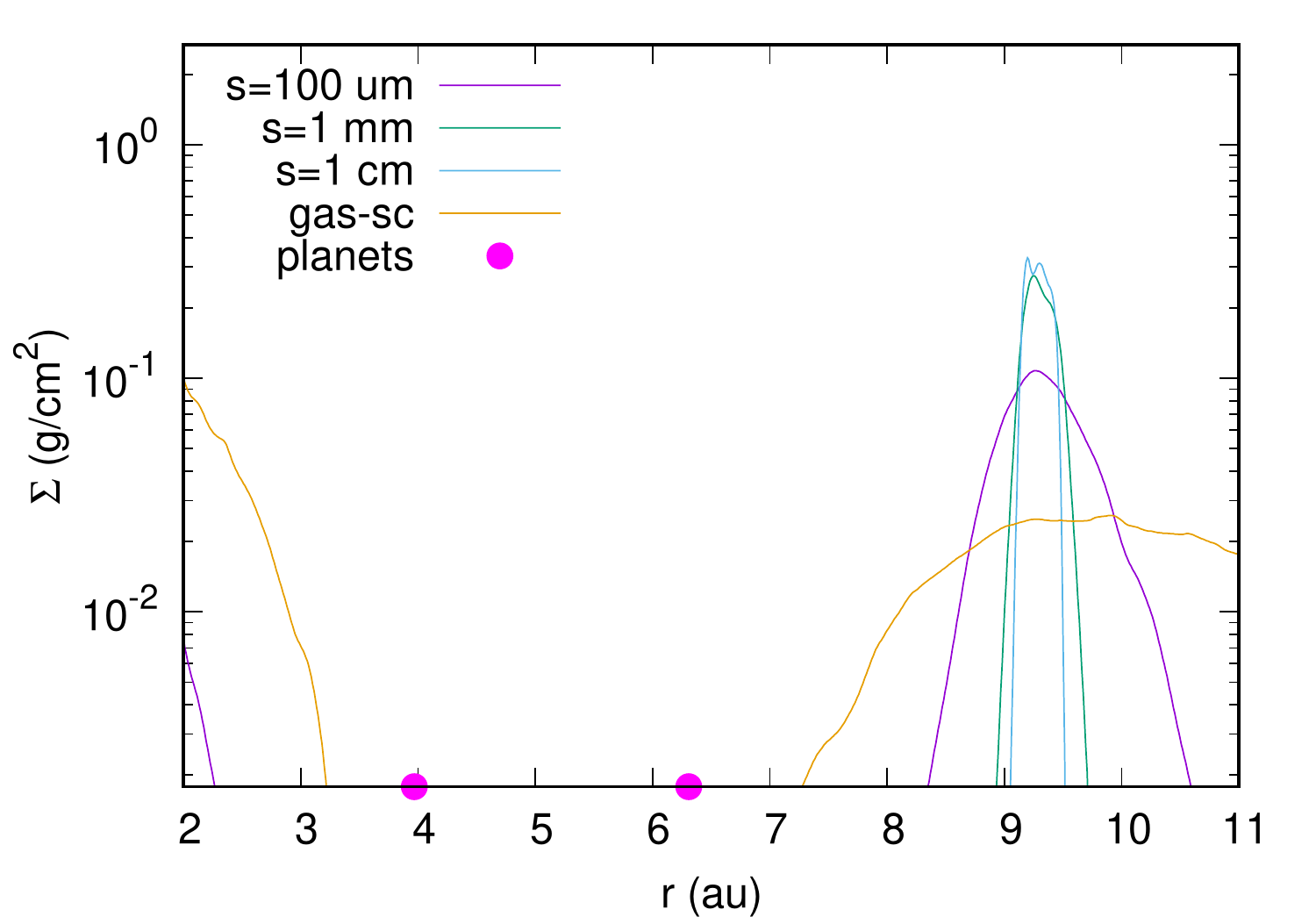}
\caption{Density profile of dust particles of different sizes ranging
from $100$~$\mu$m to $1$~cm with the planets in a 2:1 resonance. These profiles are compared to the re-scaled
gas density (multiplied by $0.0033$). The top panel is for 
a kinematic viscosity equal to $\nu = 10^{-6}$~$r^{2}_{0}\Omega_{0}$ and 
the middle panel for $\nu = 10^{-7}$~$r^{2}_{0}\Omega_{0}$.
The bottom panel refers to a model with $\nu = 10^{-6}$~$r^{2}_{0}\Omega_{0}$,
but without diffusion and back-reaction.
}
\label{fig:prof_nu_21}
\end{figure}

The dust distributions are shown in Figure~\ref{fig:prof_nu_21} 
for the two different values of kinematic viscosity and, in 
the bottom panel, for a model without the inclusion of diffusion 
and back-reaction.
Because of the significantly different migration speed, we only
compare the dust density distributions at the end of the simulation
(when the planet are located at different orbital radii).
For this resonance too a significant dust enhancement develops at 
the outer edge of the gas gap,
where dust-to-gas mass ratios can achieve values of order unity.
The inner cavity is larger at all grain sizes, compared to that produced
by the 3:2 resonant configuration.
This is possibly related to the slower outward migration rate of the planets
in the 2:1 resonance, which can reduce filtering of the dust through 
the orbits of the planets and toward the star.
Additionally, the lower gas density increases the Stokes numbers of 
the grains, reducing their drift timescale and facilitating grain
removal from the inner disc.

Comparing the top and bottom panels of Figure~\ref{fig:prof_nu_21},
one can notice significant differences in the dusty features exterior
to the planets orbits. 
Since the gas density is also different in
the two models (at those times), it is unclear how much of the
difference is caused by the action of diffusion and back-reaction
of the solids.

\begin{figure}
\includegraphics[width=\columnwidth,trim={0mm 8mm 0mm 0mm},clip]{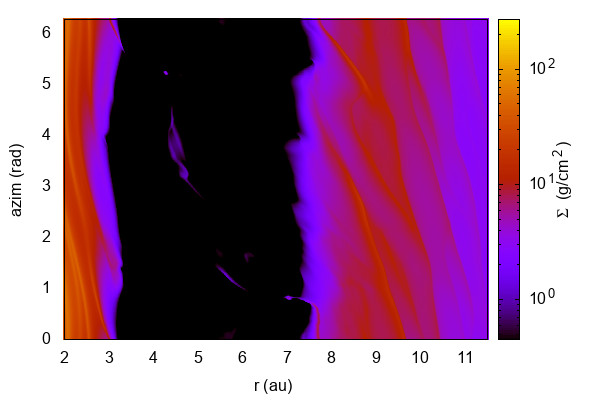}
\includegraphics[width=\columnwidth,trim={0mm 8mm 0mm 0mm},clip]{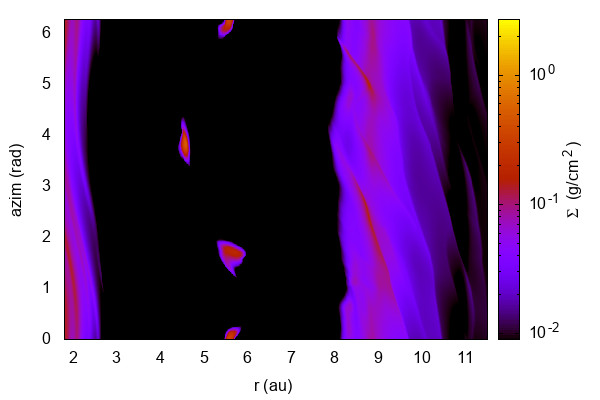}
\includegraphics[width=\columnwidth,trim={0mm 8mm 0mm 0mm},clip]{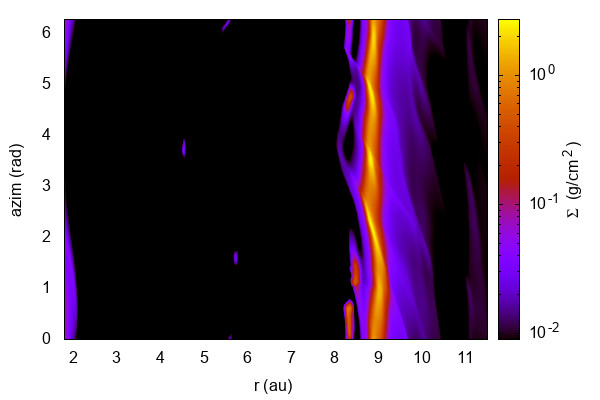}
\includegraphics[width=\columnwidth,trim={0mm 0mm 0mm 0mm},clip]{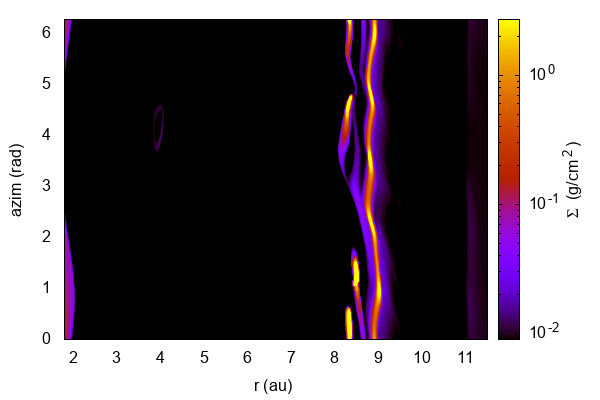}
\caption{Density maps illustrating gas and dust distributions around planets
locked in the 2:1 orbital resonance 
($\nu = 10^{-7}$~$r^{2}_{0}\Omega_{0}$)
.
The top panel refers to the surface density of the gas. The other panels
refer, respectively, to the distributions of $100$~$\mu$m, $1$~mm and 
$1$~cm dust particles. 
        }
\label{fig:map21}
\end{figure}

The orbital eccentricity of the planets can drive asymmetries in
the distribution of the disc material. Consequently, both the gas gap
outer edge and dusty rings can become asymmetric (see Figure~\ref{fig:map21}).
We did not perform a detailed analysis of ring asymmetries. However,
one possible explanation for the more asymmetric features arising
from the planets in the 2:1 resonance (compared to those in the 3:2
resonance, see Figure~\ref{fig:map32}), may be the larger eccentric
perturbation driven by the inner (more massive) planet, which tends
to have a larger orbital eccentricity in the 2:1 resonance than in 
the 3:2 resonance (the outer planets can have comparable eccentricities).

In Figure~\ref{fig:map21}, the dust density distributions are shown
for grains of various sizes
for the case with lowest viscosity,
$\nu = 10^{-7}$~$r^{2}_{0}\Omega_{0}$. 
As mentioned, the width of the gap
is significantly larger compared to the case of the 3:2 resonance
and the dusty ring at the outer edge of the gas gap is evident for
all particles, more than in the 3:2 resonance. 
For the 2:1 resonance, the effects of the dust trap appear more
marked, both at the inner and at the outer edge of the gas gap.
It is expected that after sufficient time from the beginning of
the outward migration, the disc reduces to a single overdense ring
at the outer edge of the gap, close to the outer planet orbit.
As discussed in the previous section, this feature would result
from the lack of dust supply from larger orbital radii, beyond
the outer boundary of the grid.

\section{Comparison with Alpha Viscosity}
\label{sec:AV}

To test the robustness of our results, we performed two additional 
simulations, for the 2:1 resonance, in which the kinematic viscosity
is set as $\nu=\alpha H^{2}\Omega$, where $H$ is the local pressure
scale height of the disc and the parameter $\alpha$ is assumed to 
be a constant.
Since we work with a local-isothermal disc and a constant aspect ratio,
$H\propto r$ and $\nu\propto \alpha\sqrt{r}$.
In the first model, we set $\alpha = 10^{-3}$ whereas we set 
$\alpha=10^{-4}$ in the second. 
These parameters result in $\nu\approx 10^{-6}$ and
$\approx 10^{-7}$~$r^{2}_{0}\Omega_{0}$ around the middle radius
of the computational domain.

\begin{figure}
\includegraphics[width=0.75\columnwidth, angle=-90]{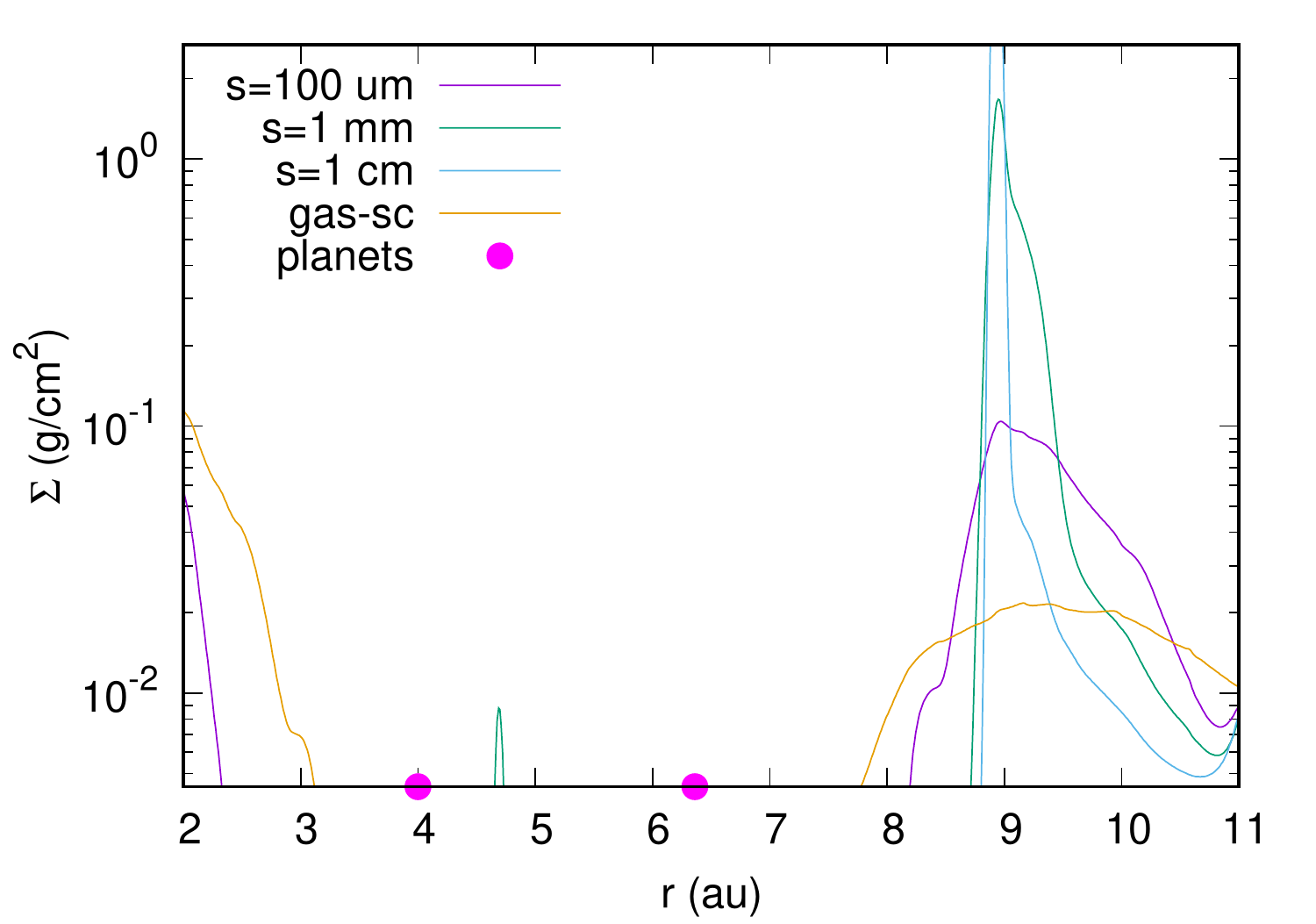}
\includegraphics[width=0.75\columnwidth, angle=-90]{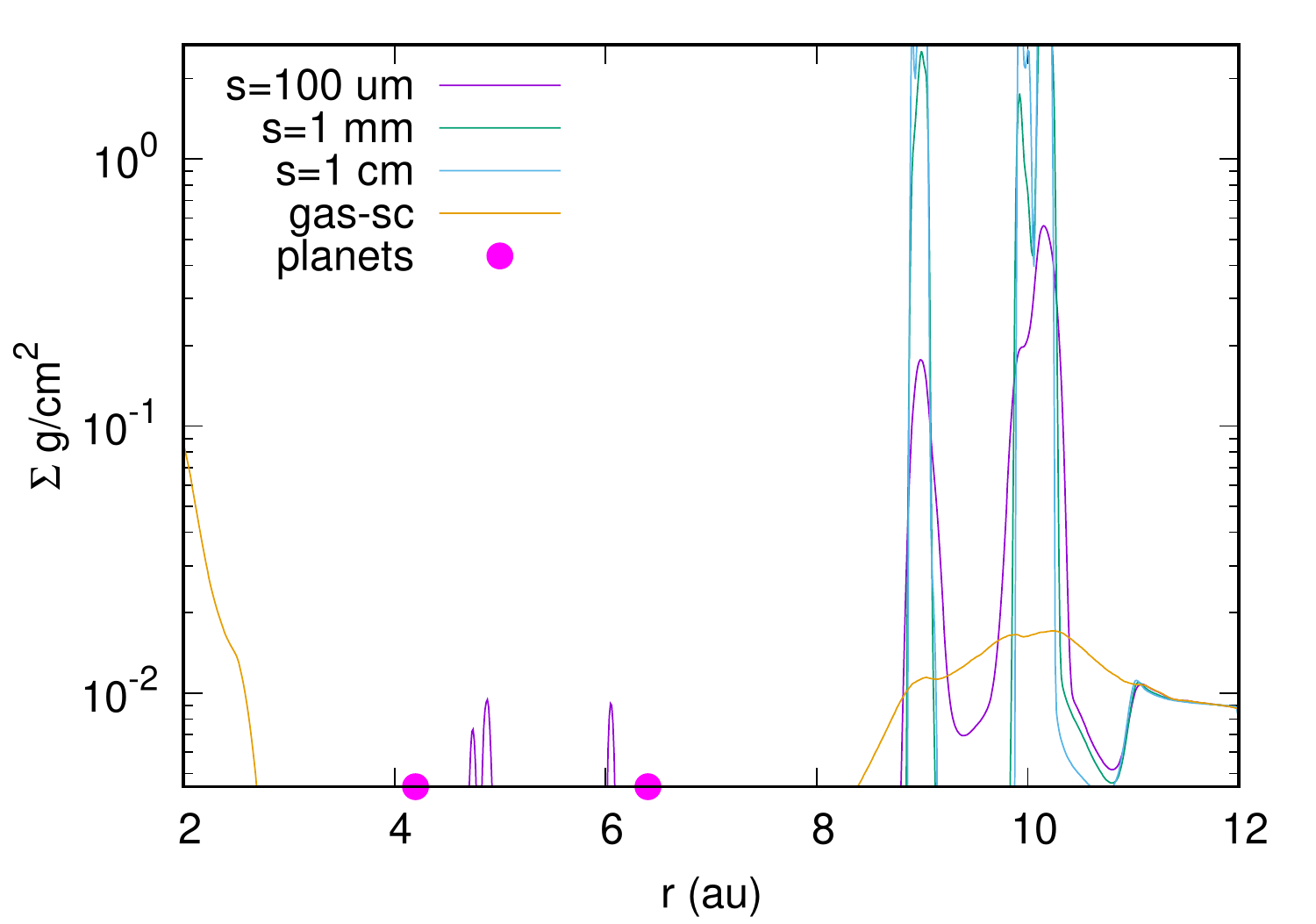}
\caption{Dust and gas density profiles 
(as in Figure~\ref{fig:prof_nu_21}) around a pair of planets locked
in a 2:1 orbital resonance, after $4 \times 10^4$ yrs, in
an $\alpha$-viscosity disc.
In the top panel $\alpha = 10^{-3}$ and in the bottom panel
$\alpha = 10^{-4}$.
}
\label{fig:prof21_alpha}
\end{figure}

The outcomes of these simulations are illustrated in  
Figure~\ref{fig:prof21_alpha}, after $4 \times 10^4$~yrs of evolution.
The density peak at the outer edge of the gas gap is clearly visible
in both cases, although there are differences in over-density morphology.
In the top panel, at $\alpha = 10^{-3}$, the peak appears similar
at all sizes while, in the bottom panel ($\alpha = 10^{-4}$), 
the peak is split in two for the largest particles ($s= 1$~mm and 
$s= 1$~cm) and composed of three separate maxima for $s= 100\,\mu$m. 
This behaviour was not observed for the 3:2 resonance but it was
already present in the simulation involving the 2:1 mean-motion
resonance obtained with a constant kinematic viscosity (see 
Figure~\ref{fig:prof_nu_21}, top panel).
A possible interpretation is that for the 2:1 resonance multiple 
dust traps develop at the outer border of the gap. 
By comparing the gas density distribution in the two cases shown
in Figure~\ref{fig:prof21_alpha}, this hypothesis appears to be 
confirmed since, in the case with $\alpha = 10^{-4}$, $\Sigma$
at the outer border of the gas gap appears more variable compared
to that of the case with $\alpha = 10^{-3}$.

The splitting of the peak in various maxima can be also observed
in the simulations of \cite{marzaridangelo2019}, for the case 
involving a pair of planet locked in the 2:1 resonance,
even if their physical model (radiative disc), the initial parameters
for the gas density and the viscosity values are different.

\section{Conclusions}
\label{sec:Con}

According to the models presented by \cite{marzaridangelo2019},
two planets in the mass range of Jupiter and Saturn, embedded 
in a circumstellar disc and migrating outwards, can strongly
affect the dust distribution in their surroundings.
They can create a transition disc with an inner cavity in the dust
distribution that expands following the outward migration of
the planets. They can also build a strong peak in the dust 
density at the outer edge of the gas gap carved by the resonant
planets, which would appear as a bright dust ring.
However, these findings may be altered by the dust diffusion
due to gas turbulence and by a change in gap morphology due 
to the back-reaction of the dust on the gas.
It is, therefore, important to test if these two processes 
weaken the dust trap at the outer edge of the gap, allowing
the dust to filter through the gap and preventing the formation
of an inner dust cavity and of the outer over-density ring. 

We used the code FARGO3D \citep{llambay2019}, in which dust
species are treated as additional pressureless fluids, and
performed a set of local-isothermal, high resolution simulations
involving giant planets locked in the 3:2 and 2:1 orbital
resonances and including the effects of diffusion and dust
back-reaction.
For the 3:2 resonance, the diffusion and back-reaction slightly
affect the dust distribution by reducing the height of the overdense
region at the outer edge of the gas gap but without removing it.
This has been tested for different values of the disc viscosity,
which determines the amount of diffusion.
This outcome proves that the dust trap is still strong enough
to halt the inward drift of the dust and to lead to the formation
of an inner cavity in the dust distribution and an overdense ring
at the outer edge of the gap. 
The details of peak heights can also be affected by continued supply
of solids from large radial distances and, therefore, depend on
boundary effects and evolution timescales.
Given the limited radial extent of the models presented herein, 
the possible feedback of very dense rings of solids on the gas
distribution was not investigated.

The 2:1 resonance also provides a robust mechanism capable of creating
an efficient dust trap. The height of the outer peak (outside the planets'
orbits) is not much affected by diffusion. Its morphology, however, appears 
more complex than it does in the 3:2 resonance situation, since in some
cases two dust peaks form at the outer edge of the gas gap, possibly due
to the development of multiple dust traps.
The robustness of these results for this resonance were tested 
by performing two additional simulations in which we switched from
a constant kinematic viscosity, $\nu$, to a constant $\alpha$ viscosity
parameter. 
In terms of large-scale features, the outcome of these simulations 
do not significantly differ from those with constant $\nu$, showing 
that the formation of the inner dust cavity and the outer peak(s) are 
not due to the viscosity parametrization (although details can depend 
on the type of viscosity).

In the models presented herein, the gas distribution interior to 
the planets' orbits is not significantly depleted. However, ongoing 
accretion on the planets, neglected here, is expected to reduce 
the gas mass flux toward the inner disc \citep{lubow2006}, possibly
leading to the formation of an inner cavity in the gas distribution
as well.
\section*{Acknowledgements}

We thank the reviewer, Cl\'ement Baruteau, whose comments helped us improve this paper.
GD acknowledges support provided by NASA’s Research
Opportunities in Space and Earth Science.
Computational resources supporting this work were provided by 
the NASA High-End Computing (HEC) Program through the NASA Advanced 
Supercomputing (NAS) Division at Ames Research Center.

\section*{Data Availability}

The data underlying the research results described in the article will 
be shared upon reasonable request to the authors.


\bibliographystyle{mnras}
\bibliography{local} 
\label{lastpage}
\end{document}